\documentstyle[preprint,prl,aps]{revtex}
\def\ypbco#1 #2{$\rm Y_{#1}Pr_{#2}BCO$}
\def\YPBCO#1 #2{$\rm Y_{#1}Pr_{#2}Ba_2Cu_3O_7$}
\begin{document}

\title{ Anomalous Behavior of the Complex Conductivity of \\
$\rm \bf Y_{1-x}Pr_xBa_2Cu_3O_7$
Observed with THz Spectroscopy }
\author{R. Buhleier\cite{RB}, S.D. Brorson\cite{SDB}, I.E. Trofimov\cite{IET},
\\ J. O. White\cite{JOW}, H.-U. Habermeier, and J. Kuhl}
\address{Max-Planck-Institut f\"ur Festk\"orperforschung, D-70569
Stuttgart, Germany}
\maketitle
\vskip .5in

\begin{abstract}
We have measured the electrodynamic properties of \YPBCO
{1-x} x\  single crystal thin films as a function of temperature using
coherent THz-time-domain spectroscopy. We obtain directly the complex
conductivity $\sigma=\sigma_1+i \sigma_2$,
the London penetration depth $\lambda_L$, the plasma frequency $\omega_p$,
and the quasiparticle scattering rate $1/\tau$.
We find that $1/\tau$ drops exponentially rapidly with $T$ below the
critical temperature in {\em all} the superconducting
samples, implying that this behavior is a {\em signature} of high-$T_c$
superconductivity.  The plasma frequency decreases with increasing Pr
content, providing evidence that Pr depletes carriers, leaving the CuO
planes {\em underdoped}. Both the conductivity in the THz region and the dc
 resistivity yield evidence for the opening of a spin gap {\em above}
$T_c$.
\end{abstract}

\newpage

The large body of vital spectroscopic data on the high-$T_c$
superconducting compound  $\rm Y_1Ba_2Cu_3O_{7-\delta}$ (YBCO)
measured in the far infrared (FIR) as well as the GHz-region
has revealed many interesting features.  Among them are: the presence  of
one\cite{rotter91} or more\cite{genzel93} energy gaps in the
superconducting state,  a temperature dependent penetration depth that is
indicative of  d-wave pairing\cite{hardy93}, and a quasiparticle
scattering rate which is proportional to frequency in the normal
state\cite{rotter91} and which drops by four orders of magnitude at
$T_c$\cite{bonn93}.

The new technique of THz-time-domain spectroscopy {\em bridges} the
gap between the FIR and the GHz regime.  THz measurements of YBCO
display a peak in the temperature dependence of $\sigma_1$ below
$T_c$, which appears to be related to a
temperature-dependent scattering rate $1/\tau$ rather than to  BCS
coherence effects\cite{nuss91}.  Other experiments using the same
technique have revealed that the penetration depth  $\lambda_L$, has a
temperature-dependence  inconsistent with BCS theory\cite{brorson94}.

Here we report results of a THz investigation into the \YPBCO {1-x} x\
(YPrBCO) system.  These measurements are of great  interest because samples
covering a wide range of critical temperatures can be obtained simply by
varying the Y/Pr ratio\cite{radousky92}.  Also, because YPrBCO is fully
stoichiometric in the oxygen content, it is more homogeneous than
oxygen-depleted YBCO and does not change from an orthorhombic to a
tetragonal lattice upon decreasing $T_c$.  Several mechanisms have been
proposed for the $T_c$ suppression, including a magnetic pair breaking
effect \cite{guo90} and depletion of holes in the CuO$_2$
planes\cite{fehrenbacher93}.

Recently, the opening  of a spin gap at temperatures above
$T_c$ in oxygen-deficient YBCO has been deduced from various
experiments e.g. dc resistivity and Hall coefficient
\cite{bucher93,ito93}, neutron scattering\cite{rossat91}, and NMR
measurements\cite{alloul89}. However, until now only one
electro-magnetic  measurement on underdoped YBCO has been linked to the
presence of a spin gap\cite{rotter91}.
In this paper, we report evidence for a spin gap
obtained from measurements {\em both} at dc {\em and} at THz frequencies in
YPrBCO, which is consistent with Pr depleting holes
from the CuO planes\cite{fehrenbacher93}.

Our spectroscopic technique involves a coherent time-domain measurement of a ps
microwave impulse $E(t)$ transmitted through the sample\cite{grischkowsky90}.
A Fourier transform yields the complex
transmission spectrum $t(\omega)$ and the complex conductivity
$\sigma(\omega)$, without the use of Kramers-Kronig analysis.
The London penetration depth $\lambda_L(T)$,
the plasma frequency $\omega_p$ in the clean limit, and the scattering time
$\tau(T)$ are obtained with the aid of a two-fluid model.

The microwave source is a biased 30~$\mu$m transmitting antenna
fabricated on low temperature grown GaAs, triggered with $\sim$100~fs
optical pulses from a colliding-pulse modelocked dye laser.
The emitted microwave pulses have spectral components spanning
the 0.1 -- 1.0 THz spectral region which is difficult to access
with conventional electronics\cite{volkov89}.
The receiver is a 30~$\mu$m antenna, fabricated on ion-implanted
silicon-on-sapphire and gated with a second pulse from the laser.
The receiver photocurrent is proportional to the incident microwave field.

We investigated five YPrBCO samples having Pr composition
$x$ = 0.0, 0.2, 0.3, 0.4, and 1.0.
The films are grown by pulsed laser deposition \cite{habermeier91} onto
NdGaO$_3$ substrates.
The film thicknesses are approximately 150~nm (Table I).
NdGaO$_3$ is the ideal substrate because it remains transparent and
nondispersive over the entire spectral bandwidth of our pulses, as
well as over the entire range of temperatures investigated here.
The excellent lattice match of NdGaO$_3$ to YPrBCO is a prerequisite
for a low density of misfit dislocations on the film-substrate
interfaces. The substrates have a (001) orientation, yielding ${\bf
{\hat c}}$ oriented, twinned films.
The critical temperatures $T_c^{\rm dc}$ (Table I) are determined
from the temperature dependent resistivity, as measured
with a four-point probe.

To calculate the conductivity, we make use of the multiple reflection
formula for the field transmitted through a layer of (complex) index
$n_2$, thickness $d$, bounded by media of index $n_1$ and $n_3$:
\begin{equation}
t(\omega) =
{{t_{12}t_{23}\exp(in_2(\omega/c)d)} \over {1 + r_{12}r_{23}\exp(2 i
n_2(\omega/c)d)}}
\label{f-p}
\end{equation}
where $t_{ij} = 2n_i/(n_i + n_j)$ is the field transmission
coefficient at the $ij$th interface,
and $r_{ij} = (n_i - n_j)/(n_i + n_j)$ is the field
reflection coefficient.
In our geometry, $n_1$ represents vacuum,
$n_2 = \sqrt{1 + i \sigma(\omega)/(\epsilon_0 \omega)}$
is the index of the superconducting layer,
and $n_3$ is the measured index of the substrate.
As $n_2 (\omega/c) d \ll 1$, and
$n_2 \gg n_3 > 1$ in our samples, Eq. (\ref{f-p}) reduces to
\begin{equation}
\label{transmission}
t(\omega) = {{1+n_3} \over {1+n_3 + Z_o \sigma(\omega) d }}
\end{equation}
where $Z_0$ is the impedance of free space.

The effect of varying Pr content $x$ (hereafter [Pr]) on the
conductivity spectra $\sigma(\omega)$ at $T=50$~K is similar to
the effect of varying the temperature for a given alloy
(Fig.~\ref{sigma_nu}). The addition of Pr has at least two interrelated
effects: a) The suppression of $T_c$ changes the partitioning between normal
and superconducting carriers. b) The total number of
carriers $N$ (or their mobility) may be reduced.
To the extent that the superconducting carriers make the largest contribution
to $\sigma_2$, both factors a) and b) lead one to expect that at a given
temperature, pure YBCO would have the largest $\sigma_2$, and it does.
Samples with 20\% and 30\% Pr have smaller values of $\sigma_2$, since
they are only slightly below their $T_c$.
The sample with 40\% Pr has $\sigma_2\approx 0$
because it is above $T_c$  at 50~K.

Only normal carriers  contribute to $\sigma_1$ for $\omega \neq 0$, but
now factors a) and b) compete.
At 50~K, $\sigma_1$ decreases with [Pr], therefore
the effect of a reduction in $N$ dominates the effect of the shift in
$T_c$ which increases the fraction of normal carriers.
The data for pure YBCO and 20\% Pr have been fit to a frequency-dependent
Drude form, and  one can see that $1/\tau$ lies in our spectral range.
For 30\% and 40\% Pr, we observe frequency independent $\sigma_1$
and can conclude that $1/\tau$ is large compared to 1~THz.
Pure PrBCO is a dielectric at 50~K, as seen by a conductivity
proportional to frequency, i.e. a dielectric constant independent of
frequency.

Examining $\sigma_2$ at a fixed frequency, e.g. 480~GHz,
we see that it is close to zero at high temperature,
but rises sharply at the onset of superconductivity,
thus providing an independent ac measurement of $T_c$ (Table I).

In all of the superconducting alloys, below $T_c$, $\sigma_1$ displays
a peak, which has been previously observed only in
fully oxygenated YBCO using microwave techniques\cite{nuss91,bonn92}
as well as measuring the thermal conductivity\cite{yu92}.
It has been attributed to a sharp rise in the scattering time $\tau$ of
normal carriers below $T_c$ offsetting the decrease in the fraction
of normal carriers, although coherence effects remain a
possibility.
For pure YBCO, the peak value is about 20 times higher than
$\sigma_1$(100~K).
With increasing [Pr] the peak height in $\sigma_1$ decreases but the
temperature corresponding to the
peak position does not shift significantly.

The normal state behavior of our samples is particularly interesting
because {\em underdoped} materials such as
(124)YBCO and oxygen deprived (123)YBCO undergo
a phase transition associated with the opening of a spin gap
at a temperature $T_D>T_c$.
Evidence for the presence of a spin gap
has been seen in neutron scattering \cite{rossat91,gehring91}, as
well as in dc resistivity measurements\cite{bucher93,ito93}.
Recent experimental \cite{takenaka92} and theoretical
\cite{fehrenbacher93} work confirms that YPrBCO alloys are also
underdoped, i.e. superconductivity is suppressed,
because holes are removed from the CuO$_2$ planes.

If the normal carriers couple strongly to spin
fluctuations, the opening of a spin gap should be accompanied by an
{\em increase} in the scattering time $\tau$, giving rise to an
{\em enhancement} in $\sigma_1$ below $T_D$ for $\omega<1/\tau$.
For pure (optimally doped) YBCO, at 480~GHz, $\sigma_1$ shows only a
single transition at $T_c$ (Fig.~\ref{sigma_T}b).
For the (underdoped) alloys,
$\sigma_1$ has two transitions, one at $T_c$, the other at a
higher temperature which increases with [Pr].
To accentuate the two transitions, Fig.~\ref{sigma_T}b is shaded in the
region bounded by $T_c$, the experimental curve, and a dashed line
representing $1 / (\alpha+\beta T)$ behavior.

The higher transition temperature seen at 480~GHz matches that of a
transition also observed in the dc resistivity (Fig.~\ref{rho_T}a).
The transition is manifested as a deviation from a linear $T$
dependence.
To show the deviation more clearly, the resistivity is first fit
to a line $\rho=\alpha+\beta T$ between 250 and 300~K, and then the
experimental values are normalized to the value
determined by the line as follows:
\begin{equation}
\rho^* = {\rho \over \alpha+\beta T}
\end{equation}
The normalized resistivity (Fig.~\ref{rho_T}b) of the YPrBCO alloys
clearly shows a transition occurring
at a temperature above $T_c$ similar to that which has been
observed in underdoped YBCO, and connected to
the opening of a spin gap\cite{bucher93,ito93}.

For the evaluation of $\lambda_L$ we are using a two fluid model
of the form:
\begin{equation}
\sigma(\omega) = {\epsilon_0 \omega_p^2 \tau
\over {1-i\omega\tau}}x_n +
{1 \over {\mu_0\lambda_L^2}}\left(- \pi\delta(\omega) + {i \over \omega}
\right) x_s,
\label{2fluid}
\end{equation}
where $\omega_p$ is the plasma frequency,
and $x_n$ and $x_s$ are the fractions of normal and
superconducting carriers, given by $x_n+x_s = 1$.
We derive the penetration depth from a fit to Eq. (\ref{2fluid}).
In our frequency range, we can ignore the Drude contribution to
$\sigma_2$ for temperatures which are more than a degree below $T_c$.
As $x_n = 0$ at $T=0$ in the two-fluid scenario, we have
$x_s = \left(\lambda_L(0)/\lambda_L(T)\right)^2$.
Shown in Fig.~\ref{lambda_vs_t} is the measured
$\left(\lambda_L(0)/\lambda_L(T)\right)^2$ {\em vs.} $T/T_c^{\rm ac}$
for all samples.   Also plotted are theoretical curves from
weak-coupling BCS theory \cite{muhlschlegel59} and
the functional form
\begin{equation}
\left(\lambda_L(0) /
\lambda_L(T) \right)^2 = 1 - (T/T_c)^{\alpha}.
\label{penetration}
\end{equation}
Power-law behavior in
the dependence of $\lambda_L$ on $T$ for $T \rightarrow 0$
is one signature of ``unconventional superconductivity'', i.e. nodes in
the energy gap in $k$-space \cite{sigrist91}.
Our data are intermediate between BCS theory and $\alpha = 2$, which is
the exponent predicted for $d$-wave pairing with impurity scattering
\cite{monthoux93}.
Microwave measurements at lower frequencies have indicated
both $\alpha = 1$ (in the $T \rightarrow 0$ limit) \cite{hardy93,ma93}
and $\alpha = 2$\cite{bonn93}.
The Gorter-Casimir two-fluid model ($\alpha = 4$) \cite{casimir34}, as
well as $d$-wave pairing ($\alpha = 1$)  are clearly inconsistent with our
data.

$\lambda_L(0)$ is obtained by extrapolating $\lambda_L(T)$ to
$T=0$ using a quadratic regression (Table I).
Our value for pure YBCO, $168 \pm 2$~nm, is close to the 145~nm value
typical for good samples\cite{bonn93}.
$\lambda_L(0)$ increases with increasing [Pr]
as the samples become less superconducting.

The plasma frequency $\omega_p$ (Table~1) is determined with
the relation $\omega_p = c/\lambda_L(0)$, which is valid in the
clean limit, provided all carriers are superconducting at $T=0$.
The clean limit assumption becomes less valid upon substitution of Pr,
therefore the values for the alloys are only estimates.
Our value of 9500~cm$^{-1}$ in pure YBCO
is close to the 12000~cm$^{-1}$ value indicative of good
sample quality\cite{genzel93}.

The quasiparticle scattering rate is calculated by fitting Eq. (\ref{2fluid})
to the data for $\sigma_1$ and using, for $x_n$, the relation
$x_n = 1-x_s = 1 - (\lambda_L(0) / \lambda_L(T) )^2$.
We find that the scattering rate $1/\tau$
decreases slowly with decreasing temperature
until $T \approx T_c$, after which it decreases exponentially
rapidly (Fig.~\ref{tau_vs_t}), possibly reflecting the opening up of a
gap in the fluctuation spectrum.
Similar behavior has been reported in YBCO samples\cite{bonn93}.
The scattering rate $1/\tau$ reveals an exponential decay below $T_c$
by more than two orders of magnitude for all samples with different [Pr]
\cite{oklein94}.
Thus we are led to conclude that this behavior is
a universal feature of high-$T_c$ superconductivity.

For $T>T_c$, we find the surprising result that $1/\tau$ decreases
with [Pr] implying that alloy scattering plays a negligible role, and
that depletion of carriers in the CuO$_2$ planes dominates the scattering.
It should be noted, however, that the drop in $1/\tau$ with [Pr]
may be due to an underestimate of $\omega_p$, if the
clean limit assumption breaks down for the alloys.
A more precise determination of the scattering rate dependence on [Pr]
above $T_c$ is beyond the scope of this letter.

In conclusion, we have presented complex conductivity experiments in the
THz range on thin films of \YPBCO{1-x} x.
$\sigma_1(T)$ reveals the anomalous coherence peak for all superconducting
alloys.  Both the conductivity in the THz region
and dc resistivity measurements provide evidence for
the opening of a spin gap in the excitation spectrum of the
underdoped samples.
$\sigma_2(\omega,T)$ yields directly the penetration depth.
For the superconducting YPrBCO alloys the plasma
frequency $\omega_p$ in the clean limit decreases with [Pr] above $T_c$
due to a reduced number of carriers.
The temperature dependent quasiparticle scattering rate $1/\tau(T)$
shows an exponential drop below $T_c$ in all the alloys.
These observations fit into the picture of Pr suppressing superconductivity
by depletion of the carrier concentration
in the superconducting CuO$_2$ planes.

We acknowledge discussions with L. Genzel, A. Bussmann-Holder,
P. Horsch, I. Mazin, T. Timusk, J.P. Carbotte, P. Littlewood,
and S. Anlage.
S.D.B. and J.O.W. acknowledge support by
the Alexander von Humboldt Foundation.

\newpage

\vskip .2in
\narrowtext
\begin{figure} 
\caption{a) $\sigma_1(\nu)$ and
b) $\sigma_2(\nu)$ for \YPBCO{1-x} x\ at $T=50$~K,
$x$ = 0 (squares), 0.2 (circles), 0.3 (dotted curve), 0.4 (dash-dotted curve),
and 1.0 (short-dashed curve).
The solid and dashed upper curves are fits to a frequency-dependent Drude
form (see text). }
\label{sigma_nu}
\end{figure}

\vskip .2in
\begin{figure} 
\caption{a) $\sigma_2(T)$ and b) $\sigma_1(T)$
for the four Pr concentrations $x = 0$ (squares),
0.2 (circles), 0.3 (solid triangles),
0.4 (open triangles) at 480 GHz.
The dashed curves in b) show $1/\rho^*$.
The vertical lines indicate $T_c^{ac}$.
For clarity, the data for $x=0.3$ in b) has been multiplied by 1.5.
}
\label{sigma_T}
\end{figure}

\vskip .2in
\begin{figure} 
\caption{a) The dc resistivity $\rho(T)$ and
b) the normalized dc resistivity $\rho^*(T)$
for $x$= 0, 0.2, 0.3, and 0.4
(solid, dashed, dotted, and dash-dotted curves).
}
\label{rho_T}
\end{figure}

\vskip .1in
\noindent
\begin{figure} 
\caption{ Normalized London penetration depth
$\lambda_L(T/T_c^{\rm ac})$
for $x$ = 0.0 (squares), 0.2 (circles), 0.3 (solid triangles),
0.4 (open triangles).
Also shown are curves corresponding to the predictions of
BCS theory (solid line),
as well as to the functional form (\protect\ref{penetration})
for $\alpha = 1$ (dash-dotted line), 2 (dashed line), and 4 (dotted line).
}
\label{lambda_vs_t}
\end{figure}

\vskip .1in
\noindent
\begin{figure} 
\caption{ Quasiparticle scattering rate $1/\tau(T)$ for
$x$ = 0 (squares), 0.2 (circles), and 0.4 (triangles) at 480~GHz.
}
\label{tau_vs_t}
\end{figure}

\vskip 1in

\begin{center}
{\bf Table I}
\vskip .2in

\begin{tabular}{|c||c|c|c|c|c|}
\hline
$x$  & $d$ (nm) & $\lambda_L(0)$ (nm) & $\omega_p$ (cm$^{-1}$) &
$T_c^{\rm dc}$ (K) & $T_c^{\rm ac}$ (K) \\
 & & & & (4-point probe) & (microwave)  \\ \hline \hline
0.0 & 155 & 168 $\pm$ 2 & 9500 & 93 & 92  \\ \hline
0.2 & 134 & 349 $\pm$ 10 & 4560 & 68 & 72 \\ \hline
0.3 & 170 & 375 $\pm$ 7 & 4244 & 53 & 59 \\ \hline
0.4 & 170 & 590 $\pm$ 15 & 2693 & 40 & 41 \\
\hline
\end{tabular}
\end{center}
\end{document}